\documentclass[12pt]{article}
\pdfoutput=1
\usepackage{epsfig}
\usepackage{amsfonts}
\usepackage{latexsym}
\usepackage{amsmath}
\usepackage{mathrsfs}
\usepackage{hyperref}
\usepackage{setspace}
\usepackage{color}
\numberwithin{equation}{section}

\begin{document}

\begin{titlepage}
\unitlength = 1mm
\begin{flushright}
KOBE-TH-14-04\\
QGASLAB-14-02
\end{flushright}

\vskip 1cm
\begin{center}

 {\Large {\textsc{\textbf{Entanglement entropy of $\alpha$-vacua 
in de Sitter space }}}}

\vspace{1.8cm}
Sugumi Kanno$^*$, Jeff Murugan$^*$,  Jonathan P. Shock$^*$ and Jiro Soda$^\dag$

\vspace{1cm}

{\it $^*$ Laboratory for Quantum Gravity \& Strings and Astrophysics, Cosmology \& Gravity Center,
Department of Mathematics \& Applied Mathematics, University of Cape Town,
Private Bag, Rondebosch 7701, South Africa}

\vspace{0.5cm}

{\it $^\dag$ Department of Physics, Kobe University, Kobe 657-8501, Japan}

\vskip 2cm

\begin{abstract}
\baselineskip=6mm
We consider the entanglement entropy of a free massive scalar field in the one parameter family of $\alpha$-vacua in de Sitter space by using a method developed by Maldacena and Pimentel. An $\alpha$-vacuum can be thought of as a state filled with particles from the point of view of the Bunch-Davies vacuum. Of all the $\alpha$-vacua we find that the entanglement entropy takes the minimal value in the Bunch-Davies solution.
We also calculate the asymptotic value of the R\'enyi entropy and find that it increases as $\alpha$ increases. We argue these features stem from pair condensation within the non-trivial $\alpha$-vacua where the pairs have an intrinsic quantum correlation. 
\end{abstract}

\vspace{1.0cm}

\end{center}
\end{titlepage}

\pagestyle{plain}
\setcounter{page}{1}
\newcounter{bean}
\baselineskip18pt

\setcounter{tocdepth}{2}

\tableofcontents

\section{Introduction}

It is well recognized that entanglement entropy is a useful tool to characterize a quantum state~\cite{Horodecki:2009zz}. 
Historically, quantum entanglement has been one of the most mysterious and fascinating features of quantum mechanics in that 
performing a local measurement may instantaneously affect the outcome of local measurements beyond the lightcone. This apparent violation of causality is known as the Einstein-Podolsky-Rosen paradox~\cite{Einstein:1935rr}. However, since information does not get transferred in such a measurement, causality remains intact. There are many phenomena which we are now finding that quantum entanglement may play a role, including bubble nucleation~\cite{Coleman:1980aw}. Schwinger pair production in a constant electric field can be considered as an analogue of bubble nucleation in a false vacuum. In the case of pair creation, electron-positron pairs are spontaneously created with a certain separation and such particle states should then be quantum correlated. Recent studies of the Schwinger effect infer that observer frames will be strongly correlated to each other when they observe the nucleation frame~\cite{Garriga:2012qp, Garriga:2013pga, Frob:2014zka}.

Entanglement entropy has now been established as a suitable measure of the degree of entanglement of a quantum system. Entanglement entropy has since become a useful tool in understanding phenomena in condensed matter physics, quantum information and high energy physics. For example, entanglement entropy plays the roll of an order parameter in condensed matter systems and thus the phase structure can be examined using this measure of quantum correlation. It would be interesting to consider the consequences of a measurable entanglement entropy in a cosmological setting, especially in view of the bubble nucleation and the multiverse. Indeed, it may be possible to investigate whether a universe entangled with our own universe exists within the multiverse framework. Such a scenario may be observable through the cosmic microwave background radiation (CMB).

To calculate the entanglement entropy in quantum field theories explicitly has, until recently, not been an easy task. In \cite{Ryu:2006bv}, Ryu and Takayanagi proposed a method of calculating the entanglement entropy of a strongly coupled quantum field theory with a gravity dual using holographic techniques. This has proved extremely powerful and their formula has passed many consistency checks~\cite{Takayanagi:2012kg}. Consequently, entanglement entropy, especially within a holographic context has been attracting a great deal of attention of late.

In~\cite{Maldacena:2012xp} Maldacena and Pimentel developed a method to explicitly calculate the entanglement entropy in a quantum field theory in the Bunch-Davies vacuum of de Sitter space and discussed the gravitational dual of this theory and its holographic interpretation. In this paper, we extend the calculation of the entanglement entropy in the Bunch-Davies vacuum to $\alpha$-vacua. The use of conformal symmetry of the de Sitter invariant Bunch-Davies vacuum as utilized by Maldacena and Pimental can be also extended to the $\alpha$-vacua and this will significantly simplify the calculation.

Our interest in $\alpha$-vacua is three-fold: Firstly, these new examples serve to further develop our understanding of the nature of entanglement entropy in the non-trivial vacua with de Sitter invariance. Second, understanding entanglement entropy in this de Sitter invariant family of backgrounds will provide a non-trivial check of the holographic methods employed by Ryu and Takayanagi which even today is the only tool at our disposal to access the entanglement entropy of strongly-coupled quantum field theories. We would like to clarify here how the change of the vacuum can be implemented into the holographic scheme. This is related to the issue of how to describe the entanglement entropy of excited states from the holographic point of view. Our final motivation stems from cosmology. In conjunction with CMB observations, 
it has been suggested~\cite{Ashoorioon:2014nta} that non-Bunch-Davies vacua are preferable to explain the results of BICEP2~\cite{Ade:2014xna}. 

In this paper we calculate the entanglement entropy of a massive scalar field in the family of $\alpha$-vacua in a fixed de Sitter background. Note that this is not the entropy of the metric on the de Sitter horizon which has already been discussed extensively in the literature~\cite{Gibbons:1977mu}\cite{Spradlin:2001pw}. Intuitively, an $\alpha$-vacuum can be thought of as a state of pair 
condensation or, alternatively, as a squeezed state~\cite{Bousso:2001mw,Danielsson:2002mb,Einhorn:2003xb,Collins:2003zv}. Thus the quantum uncertainty of the state is heavily constrained. The pair condensation itself has an intrinsic quantum correlation associated with it. 
When the rate of pair condensation increases, it is expected that the quantum correlation would increase. Hence, we expect an increase of the entanglement entropy with increasing $\alpha$ parameter, corresponding to the increased pair condensation. We are also able to investigate the R\'enyi entropy using the same mathematical techniques. This will gives rise to a new measure of quantum
 correlation in these vacua and we will explore the $\alpha$ dependence. 
We try to give a holographic interpretation in the discussion, however, we find it difficult to implement the change of the vacuum in a conventional manner.

The paper is organized as follows. In section 2, we review the method developed by Maldacena and Pimentel with some comments relevant to the calculation of the entanglement entropy of $\alpha$-vacua. In section 3, we introduce the $\alpha$-vacua and calculate the relevant density matrix and the entanglement entropy. We also evaluate the R\'enyi entropy. We conclude in section 4 with some summary remarks and speculation about a possible holographic interpretation of our results.

\section{A review of entanglement entropy of the Bunch-Davies vacuum}

In  \cite{Maldacena:2012xp}, Maldacena and Pimentel developed a method to calculate a specific contribution to the entanglement entropy of a massive scalar field in de Sitter space explicitly. They showed that the long range correlations implied by the entanglement entropy are maximal for small masses and decay exponentially as the mass increases. Here we will review the formalism developed in that paper before extending it to $\alpha$-vacua in Section \ref{sec:alpha}.

\subsection{Entanglement entropy}

The entanglement entropy is a quantity which characterizes quantum correlations of a system. In particular it is the long-range correlations in which we will be interested. It can be thought of as a measure of how much we can discover about the full state of a system, given only a subsystem of it to measure.
To explain this, let us divide the system into two subsystems $A$ and $B$. The Hilbert space becomes a direct product ${\cal H}={\cal H}_A\otimes{\cal H}_B$. As an illustration, we choose a special state
\begin{eqnarray}
|\Psi\rangle=\sum_i c_i|\,i\,\rangle_A|\,i\,\rangle_B\,,
\end{eqnarray}
where $c_i$ is the amplitude of finding the $i$-th state. In this case, the density matrix is
\begin{eqnarray}
\rho= |\Psi\rangle\langle\Psi|=\sum_i\sum_j c_i c_j^*\,|\,i\,\rangle_A|\,i\,\rangle_B\,
{}_{A}\langle\,j\,|{}_{B}\langle\,j\,|\,.
\end{eqnarray}
If we trace over the degrees of freedom of $B$, we find that the density matrix of the subsystem $A$ is given by
\begin{eqnarray}
\rho_A={\rm Tr}_B\,\rho
=\sum_{i,j,k}c_i c_j^*\, {}_{B}\langle\,k
|\,i\,\rangle_A|\,i\,\rangle_B\,
{}_{A}\langle\,j\,|{}_{B}\langle\,j\,|\,k\,\rangle_B
=\sum_k|c_k|^2\,|\,k\,\rangle_A {}_A\langle\,k\,|\,,
\end{eqnarray}
and the density matrix is normalized to 1 because of the conservation of probability
\begin{eqnarray}
{\rm Tr}_A\,\rho_A=\sum_k|c_k|^2=1\,.
\end{eqnarray}
However, ${\rm Tr}_A\,\rho_A^2=\sum_k|c_k|^4\neq1$ in general. 

The entanglement entropy is defined via the density matrix as the Von-Neumann entropy
\begin{eqnarray}
S= - {\rm Tr}_A\,\rho_A\, \log \rho_A
=-\sum_k|c_k|^2\log |c_k|^2\,,
\label{ee}
\end{eqnarray}
where we traced over the subsystem $A$. For a pure state such as $c_1=1\,, c_2=c_3=\cdots c_N=0$, the entanglement entropy is given by $S=0$. For a mixed state such as $c_1=c_2=\cdots c_N=1/\sqrt{N}$, where $N$ is the dimensionality of the correlated Hilbert space, the entanglement entropy takes the maximum value
\begin{eqnarray}
S=-\sum_k^N\frac{1}{N}\log \frac{1}{N}=\log N \,.
\end{eqnarray}
Since the number $N$ describes the extent to which the system correlates, the entanglement entropy is certainly a measure of the quantum correlations.
In other words, the entanglement entropy is related to the relevant degrees of freedom in the system. For instance, in a two dimensional conformal field theory the entanglement entropy is proportional to the central charge which counts the degrees of freedom in such a system~\cite{Holzhey:1994we}.

\subsection{Setup of entanglement entropy in de Sitter space}

In order to study entanglement entropy in $3+1$-dimensional de Sitter space we consider a closed surface $S^2$ in a hypersurface at fixed time. This divides the spacelike hypersurface into an inside region ($A$) and an outside region ($B$). The total Hilbert space, as in the previous section can then be written as a direct product ${\cal H}={\cal H}_{\rm in}\otimes{\cal H}_{\rm out}$. From this we can trace over the outside region to construct the density matrix for the internal region $\rho_{\rm in}={\rm Tr}_{\rm out}\,|\Psi\rangle\langle\Psi|$. From this we can then obtain the entanglement entropy defined in Eq.~(\ref{ee}).

In order to apply this procedure to de Sitter space, we first consider the closed surface in the flat chart. In the flat chart of de Sitter space, the metric reads
\begin{eqnarray}
ds^2 = \frac{1}{H^2 \eta^2} \left[ -d\eta^2 + \delta_{ij}\,dx^i dx^j\right]\,,
\end{eqnarray}
where indices $(i,j)$ denote the three spatial components. $H$ is the Hubble parameter and $\eta$ is conformal time.

We consider a free scalar field of mass $m$ on a $\eta={\rm constant}$ hypersurface. The entanglement entropy associated to this field in the field theory consists of UV divergent and UV finite parts 
\begin{eqnarray}
S = S_{\rm UV-div} + S_{\rm UV-fin}\,.
\end{eqnarray}
The divergent part is well known and takes the form~\cite{Bombelli:1986rw}\cite{Srednicki:1993im}
\begin{eqnarray}
S_{\rm UV-div} = c_1 \frac{\cal A}{\epsilon^2}  
+ \log (\epsilon H) \left[\, c_2 + c_3\,{\cal A}\,m^2 +c_4\,{\cal A}\,H^2\,\right]\,,
\end{eqnarray}
where $c_i$ are numerical coefficients. Here, $\epsilon$ is the UV-cutoff
and ${\cal A}$ is the proper area of the shared surface of the two regions. Since all of these terms arise in flat space and from local effects, we are not interested in this part.
 The UV-finite part contains information about the long-range correlations of the quantum state in de Sitter space. We can expect the IR behavior ($\eta\rightarrow 0$ limit) of the UV-finite part to be of the form
\begin{eqnarray}
S_{\rm UV-fin} = c_5\,{\cal A}\,H^2 - \frac{c_6}{2} \log ({\cal A}\,H^2) + {\rm finite}
 = c_5\,\frac{\cal\tilde A}{\eta^2} + c_6\log \eta + {\rm finite}\,,
\label{uvfin}
\end{eqnarray}
where ${\cal\tilde A}={\cal A}H^2\eta^2$ is the area of the surface in comoving coordinates.
The quantity we calculate to get the information about the long range correlations of the state is $c_6$, which is a cut-off independent quantity.

\begin{figure}[t]
\vspace{-2.5cm}
\includegraphics[height=11cm]{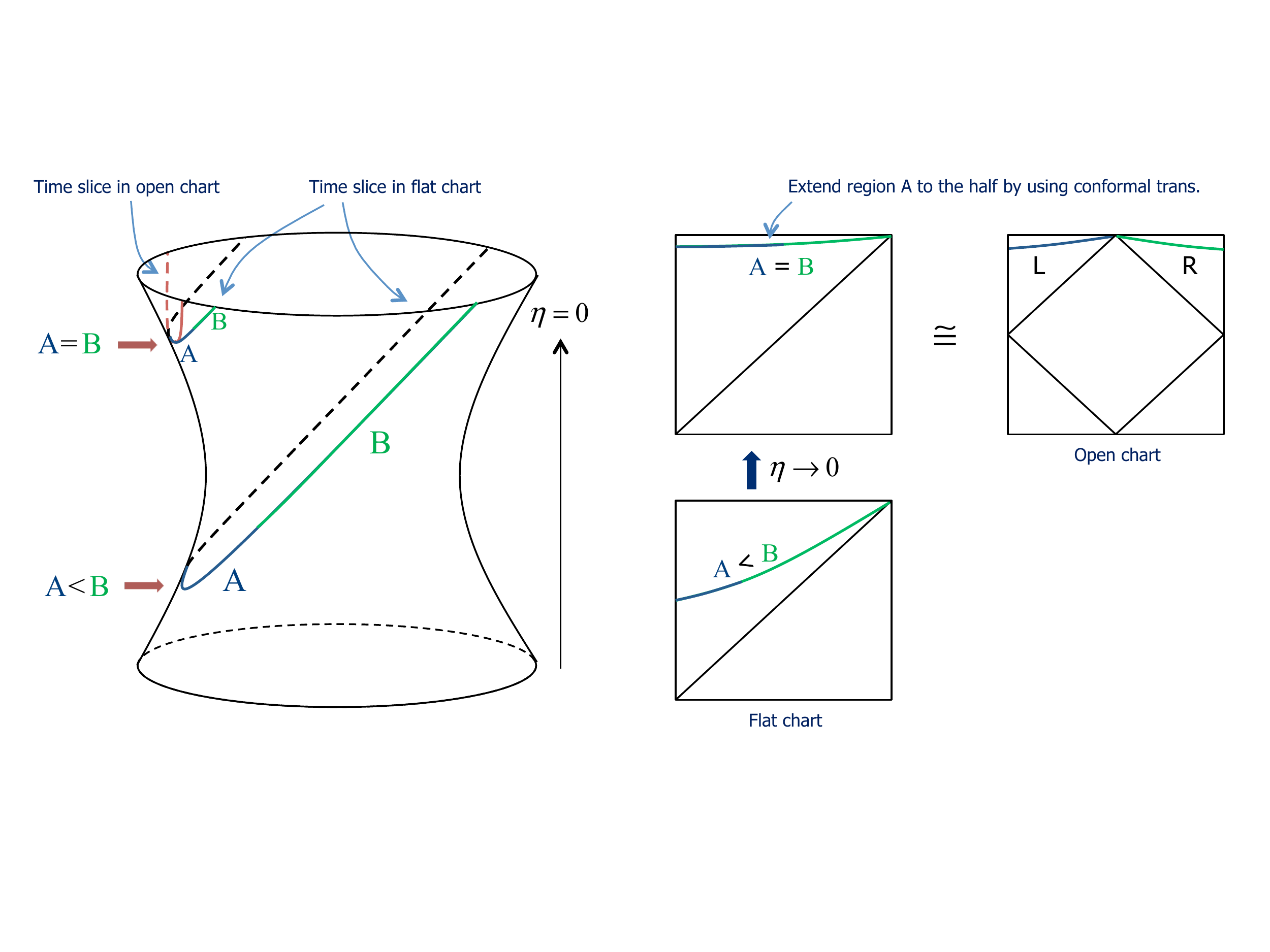}\centering
\vspace{-2cm}
\caption{De Sitter space and the Penrose diagram.}
\label{fig1}
\end{figure}

In order to compute $c_6$, we take the radius of the surface to be much bigger than the de Sitter horizon, $R\gg  R_{\rm dS}=H^{-1}$. In the asymptotic future $\eta\rightarrow 0$, we can use a conformal transformation of the de Sitter invariance to map the shared $S^2$ to the equator of the $S^3$ hypersurface at fixed time. Once we can divide the region in half in the asymptotic future, the flat chart is indistinguishable from the open chart as in Figure~\ref{fig1}. Once we can map the surface on the boundary between the $L$ and $R$ regions of the open chart in the asymptotic future, it is convenient to trace over the outside region of the surface to obtain the density matrix of the inside region. Thus, the detailed calculation will be able to be performed in the open chart as follows.

\subsection{Entanglement entropy in the Bunch-Davies vacuum}

The following section follows the derivation of \cite{Maldacena:2012xp}. The outline of the section is as follows. We will first solve the equations of motion for a free scalar field in the $L$ and $R$ regions and find the solution written in terms of the appropriate operators which annihilate the Bunch-Davies vacuum. We will then perform a transformation of this state into the operators written in terms of creation and annihilation operators on the $L$ and $R$ regions separately. This will then allow us to trace over one part of the space to obtain the density matrix and eventually the entanglement entropy. We start by finding the solutions to the equation of motion on the two sections of the space and find a suitable solution which is analytic on going between the two regions.

The open chart of de Sitter space is studied in detail in \cite{Sasaki:1994yt}. 
Now, we consider a free scalar field of mass $m$ with the canonical action given by
\begin{eqnarray}
S=\int d^4 x \sqrt{-g} \left[\, -\frac{1}{2}\,g^{\mu\nu}\partial_\mu\phi\,\partial_\nu \phi
-\frac{m^2}{2}\phi^2\,\right]\,.
\end{eqnarray}
We write the metric in each region $R$ and $L$ in terms of the local definitions of $t$ and $r$ as defined from the Euclidean coordinates
\begin{eqnarray}
ds^2_R&=&H^{-2}\left[-dt^2_R+\sinh^2t_R\left(dr^2_R+\sinh^2r_R\,d\Omega^2\right)
\right]\,,\nonumber\\
ds^2_L&=&H^{-2}\left[-dt^2_L+\sinh^2t_L\left(dr^2_L+\sinh^2r_L\,d\Omega^2\right)
\right]\,,
\end{eqnarray}
where $d\Omega^2$ is the metric on the two-sphere. These coordinate systems are obtained by analytic continuation from the Euclidean metric. Since it is natural to choose the Euclidean vacuum (the Bunch-Davies vacuum \cite{Bunch:1978yq,Chernikov:1968zm,Hartle:1983ai}) as the initial condition, we need to find the positive frequency mode functions corresponding to the Euclidean vacuum. After separation of variables 
\begin{eqnarray}
\phi=  \frac{H}{\sinh t}\,  \chi_p (t) \,Y_{p\ell m}(r,\Omega)\,,
\end{eqnarray} 
the equations of motion for $\chi_p$ and $Y_{p\ell m}$ in the $R$ or $L$ regions are found to be
\begin{eqnarray}
&&\left[\,\frac{\partial^2}{\partial t^2}+3\coth t\,\frac{\partial}{\partial t}
+\frac{1+p^2}{\sinh^2t}+\frac{m^2}{H^2}\right]\chi_p(t)=0\,,\\
&&\left[\,\frac{\partial^2}{\partial r^2}+2\coth r\,\frac{\partial}{\partial r}
-\frac{1}{\sinh^2r}\,\rm\bf L^2\,\right]Y_{p\ell m}(r,\Omega)=-(1+p^2)Y_{p\ell m}(r,\Omega)\,,
\end{eqnarray}
where $\rm\bf L^2$ is the Laplacian operator on the unit two-sphere, the $Y_{p\ell m}$ are eigenfunctions on the three-dimensional hyperboloid and the temporal and radial coordinates are left undistinguished for the $R$ and $L$ regions.
The solutions can be found explicitly in \cite{Sasaki:1994yt}.  By defining a parameter
\begin{eqnarray}
\nu=\sqrt{\frac{9}{4}-\frac{m^2}{H^2}}\,,
\end{eqnarray}
the time dependent part of $\chi_p(t)$ is given by
\begin{eqnarray}
\chi_{p,\sigma}(t)=\left\{
\begin{array}{l}
\frac{1}{2\sinh\pi p}\left(\frac{e^{\pi p}-i\sigma e^{-i\pi\nu}}{\Gamma(\nu+ip+1/2)}P_{\nu-1/2}^{ip}(\cosh t_R)
-\frac{e^{-\pi p}-i\sigma e^{-i\pi\nu}}{\Gamma(\nu-ip+1/2)}P_{\nu-1/2}^{-ip}(\cosh t_R)
\right)\,,\\
\\
\frac{\sigma}{2\sinh\pi p}\left(\frac{e^{\pi p}-i\sigma e^{-i\pi\nu}}{\Gamma(\nu+ip+1/2)}P_{\nu-1/2}^{ip}(\cosh t_L)
-\frac{e^{-\pi p}-i\sigma e^{-i\pi\nu}}{\Gamma(\nu-ip+1/2)}P_{\nu-1/2}^{-ip}(\cosh t_L)
\right)\,,\\
\end{array}
\right.
\label{sty}
\end{eqnarray}
where the index $\sigma$ takes the values $\pm 1$ and $P^{\pm ip}_{\nu-\frac{1}{2}}$ are Legendre functions. This is a solution supported on the $R$ and $L$ regions respectively. Two independent solutions for each region are distinguished by the sign of $\sigma$. Note that the solutions are obtained by analytic continuation between the $L$ and $R$ regions. This procedure requires the solutions to be analytic in the Euclidean hemisphere and produces the factor $e^{-\pi p}$ in the above solutions. In this way, the Bunch-Davies vacuum is selected. As the Bunch-Davies vacuum is de Sitter invariant, it is legitimate to use the mapping in Figure \ref{fig1}.

We expand the field in terms of the creation and anihilation operators,
\begin{eqnarray}
\hat\phi(t,r,\Omega) = \sum_{\sigma,\ell,m} \int dp 
\left[\,a_{\sigma p\ell m}\,u_{\sigma p\ell m}(t,r,\Omega)
+a_{\sigma p\ell m}^\dagger\,u^*_{\sigma p\ell m}(t,r,\Omega)\,\right]\,,
\label{phi}
\end{eqnarray}
where $a_{\sigma p\ell m}$ satisfies $a_{\sigma p\ell m}|{\rm BD}\rangle=0$. The mode function $u_{\sigma p\ell m}(t,r,\Omega)$ representing the Bunch-Davies vacuum is
\begin{eqnarray}
u_{\sigma p\ell m} = \frac{H}{\sinh t}\,
\chi_{p,\sigma}(t)\,Y_{p\ell m} (r, \Omega)\,.
\label{bdmf}
\end{eqnarray}
In order to calculate the density matrix, we write the states in a matrix form. If we write the bases of the $L$ and $R$ regions in a simple form $P^{R, L}\equiv P_{\nu-1/2}^{ip}(\cosh t_{R,L})\,, P^{*\,R, L}\equiv P_{\nu-1/2}^{-ip}(\cosh t_{R,L})$, the two lines of Eq.~(\ref{sty}) are expressed in one line
\begin{eqnarray}
\chi^{\sigma} = N_p^{-1} \sum_{q=R,L} \left[\,
 \alpha_q^\sigma\,P^q + \beta_q^\sigma\,P^{*\,q} 
\,\right]\,,
\label{sty2}
\end{eqnarray}
where the support of the functions of $t_L$ and $t_R$ are the relevant sub-regions. $N_p$ is a normalization factor including the $1/\sinh \pi p$ in Eq.~(\ref{sty}) and
\begin{eqnarray}
&&\alpha_R^\sigma = \frac{e^{\pi p} -i\sigma e^{-i\pi \nu}}{\Gamma (\nu+ip +\frac{1}{2})}\qquad,\qquad
\beta_R^\sigma =-\frac{e^{-\pi p} -i\sigma e^{-i\pi \nu}}{\Gamma (\nu-ip +\frac{1}{2})} \,\,,\\
&&\alpha_L^\sigma =\sigma\,\frac{e^{\pi p} -i\sigma e^{-i\pi \nu}}{\Gamma (\nu+ip +\frac{1}{2})}  
\quad\,,\qquad
\beta_L^\sigma =-\sigma\,\frac{e^{-\pi p} -i\sigma e^{-i\pi \nu}}{\Gamma (\nu-ip +\frac{1}{2})}   \,\,.
\end{eqnarray}
The complex conjugate of Eq.~(\ref{sty2}) which is needed in Eq.~(\ref{phi}) is
\begin{eqnarray}
\chi^{*\,\sigma}=N_p^{-1}\sum_{q=R,L} \left[\,
{\beta^{*}{}_q}^{\!\!\!\sigma}\,P^q + {\alpha^{*}{}_q}^{\!\!\!\sigma}\,P^{*\,q} 
\,\right]\,.
\end{eqnarray}
Then Eq.~(\ref{sty}) and its conjugate can be accommodated into the simple matrix form
\begin{eqnarray}
\chi^I=N_p^{-1}\,M^I{}_J\,P^J\,,
\end{eqnarray}
where the capital indices $(I,J)$ run from 1 to 4 and
\begin{eqnarray}
\chi^I=\left(\,\chi^\sigma\,,\chi^{*\,\sigma}\,\right)\,,\quad
M^I{}_J=\left(
\begin{array}{ll}
\alpha^\sigma_q & \beta^\sigma_q \vspace{3mm}\\
{\beta^{*}{}_q}^{\!\!\!\sigma} & {\alpha^{*}{}_q}^{\!\!\!\sigma} \\
\end{array}\right)\,,\quad
P^J=\left(\,P^R\,,P^L\,,P^{*\,R}\,, P^{*\,L}\,\right)\,.
\end{eqnarray}

Now we focus on the time dependent part\footnote{We omit the factor $1/\sinh t$ because it will be canceled when comparing Eqs.~(\ref{phi2}) with (\ref{phi3}).} of the field operator, which is written as
\begin{eqnarray}
\phi(t)=a_I\,\chi^I=N_p^{-1}a_I\,M^I{}_J\,P^J\,,\qquad
a_I=\left(\,a_\sigma\,,\,a_\sigma^\dagger\,\right)\,,
\label{phi2}
\end{eqnarray}
where the mode functions are defined via the appropriate annihilation of the Bunch-Davies vacuum. Note that this relation can be regarded as the Bogoliubov transformation by changing the mode functions defined in the Bunch-Davies vacuum into the Legendre functions, which realize the positive frequency modes in the past. The Bogoliubov coefficients are then expressed by $\alpha$ and $\beta$ in the matrix $M$.
To trace out the region $R$ (or $L$) in the end to obtain the density matrix of the $L$ (or $R$) regions, we need to know the relation between the Bunch-Davies vacuum and the $R$ and $L$-vacua. By introducing new creation and anihilation operators $b_J$ defined such that $b_R|R\rangle=0$ and $b_L|L\rangle=0$, we expand the field operator as
\begin{eqnarray}
\phi(t)=N_p^{-1}b_J\,P^J\,,\qquad
b_J=\left(\,b_R\,,\,b_L\,,\,b_R^\dagger\,,\, b_L^\dagger\,\right)\,.
\label{phi3}
\end{eqnarray}
Note that we took the Legendre functions as the mode functions of the $R,L$-vacua because they realize the positive frequency mode in the past. By comparing Eqs.~(\ref{phi2}) with (\ref{phi3}), we find the relation between $a_J$ and $b_J$ such as
\begin{eqnarray}
a_J=b_I\left(M^{-1}\right)^I{}_J\,,\qquad
\left(M^{-1}\right)^I{}_J=\left(
\begin{array}{ll}
\xi_{q\sigma} & \delta_{q\sigma} \vspace{3mm}\\
\delta_{q\sigma}^* & \xi_{q\sigma}^* \\
\end{array}\right)\,,\qquad
\left\{
\begin{array}{l}
\xi=
\left(\alpha-\beta\,\alpha^{*\,-1}\beta^*\right)^{-1}\,,\vspace{3mm}\\
\delta=-\alpha^{-1}\beta\,\xi^*\,.
\end{array}
\right.
\label{xidelta1}
\end{eqnarray}
This leads to the relation between $a_\sigma$ and $b_q$
\begin{eqnarray}
a_\sigma=\sum_{q=R,L}\left[\,\xi_{q\sigma}\,b_q+\delta_{q\sigma}^*\,b_q^\dagger\,\right]\,.
\label{ab}
\end{eqnarray}

Thus, the Bunch-Davies vacuum can be regarded as the Bogoliubov transformation from 
the $R,L$-vacua as
\begin{eqnarray}
|{\rm BD}\rangle = \exp\left(\frac{1}{2}\sum_{i,j=R,L}m_{ij}\,b_i^\dagger\, b_j^\dagger\right) |R\rangle|L\rangle\,,
\label{bogoliubov1}
\end{eqnarray}
where the operators $b_i$ satisfy the commutation relation $[b_i,b_j^\dagger]=\delta_{ij}$. The condition $a_\sigma|{\rm BD}\rangle=0$ gives
\begin{eqnarray}
m_{ij}=-\delta_{i\sigma}^*\left(\xi^{-1}\right)_{\sigma j}
=-\frac{\Gamma\left(\nu-ip+1/2\right)}{\Gamma\left(\nu+ip+1/2\right)}
\frac{2\,e^{i\pi\nu}}{e^{2\pi p}+e^{2i\pi\nu}}
\left(
\begin{array}{cc}
\cos \pi\nu & i\sinh p\pi \vspace{1mm}\\
i\sinh p\pi & \cos \pi\nu \\
\end{array}
\right)\,.
\label{mij1}
\end{eqnarray}
The phase terms are unimportant for $\nu^2>0$. We will comment on the case $\nu^2<0$ later in section \ref{largemass1}. Here, we consider $\nu^2>0$ and write them as $e^{i\theta}$ for simplicity. Then 
\begin{eqnarray}
m_{ij}=e^{i\theta}\frac{\sqrt{2}\,e^{-p\pi}}{\sqrt{\cosh 2\pi p+\cos 2\pi\nu}}
\left(
\begin{array}{cc}
\cos \pi\nu & i\sinh p\pi \vspace{1mm}\\
i\sinh p\pi & \cos \pi\nu \\
\end{array}
\right)\,,
\label{mij2}
\end{eqnarray}
where $e^{i\theta}$ contains all phase factors. We write $m_{RR} = m_{LL}\equiv \omega$, which is real and $m_{LR}=m_{RL}\equiv \zeta$, which is purely imaginary for positive $\nu^2$. 

It is still difficult to trace over the $R$ (or $L$) degrees of freedom when the state is written in the form of Eq.~(\ref{bogoliubov1}).
Thus, we perform the Bogoliubov transformation again by introducing new operators $c_R$ and $c_L$
\begin{eqnarray}
c_R = u\,b_R + v\,b_R^\dagger \,,\qquad\quad
c_L = \bar{u}\,b_L + \bar{v}\,b_L^\dagger\,,
\label{bc}
\end{eqnarray}
to get the relation
\begin{eqnarray}
|{\rm BD}\rangle = \exp\left(\gamma_p\,c_R^\dagger\,c_L^\dagger\,\right)|R'\rangle|L'\rangle\,,
\label{bogoliubov2}
\end{eqnarray}
where $|u|^2-|v|^2=1$ and $|\bar{u}|^2-|\bar{v}|^2=1$ are assumed. The operators satisfy the commutation relation $[c_i,c_j^\dagger]=\delta_{ij}$. It should be noted that the Bogoliubov transformation does not mix $L$ and $R$ Hilbert spaces
although the vacuum is changed by this transformation from $|R\rangle|L\rangle$ into $|R^\prime\rangle|L^\prime\rangle$. 
The consistency conditions for Eq.~(\ref{bogoliubov2}) are
\begin{eqnarray}
c_R\,|{\rm BD}\rangle= \gamma_p\,c_L^\dagger\,|{\rm BD}\rangle \,,\qquad 
c_L\,|{\rm BD}\rangle = \gamma_p\,c_R^\dagger\,|{\rm BD}\rangle\,.
\label{consistency}
\end{eqnarray}
Putting Eqs.~(\ref{bc}) and (\ref{bogoliubov2}) into Eq.~(\ref{consistency}), we find the system of four homogeneous equations
\begin{eqnarray}
&&\omega\,u + v -\gamma_p\,\zeta\,\bar{v}^* =0 \ , \qquad 
\zeta\,u - \gamma_p\,\bar{u}^* - \gamma_p\,\omega\,\bar{v}^* =0\,,
\label{system1}\\
&&\omega\,\bar{u} + \bar{v} -\gamma_p\,\zeta\,v^* =0 \ , \qquad
\zeta\,\bar{u} - \gamma_p\,u^* - \gamma_p\,\omega\,v^* =0\,.
\label{system2}
\end{eqnarray}
Here, $\omega$ is real $\omega^*=\omega$ and $\zeta$ is pure imaginary $\zeta^*=-\zeta$ for positive $\nu^2$. Taking the compex conjugate of Eq.~(\ref{system1}), we find that we can set $v^* =\bar{v}$ and $u^* =\bar{u}$ if $\gamma_p$ is pure imaginary $\gamma_p^*=-\gamma_p$. Then
Eq.~(\ref{system2}) becomes  identical with Eq.~(\ref{system1}) and the system is reduced  to that of two homogeneous equations. The normalization condition $|u|^2-|v|^2=1$ must be imposed. 

In order to have a non-trivial solution in the system of equations (\ref{system1}), $\gamma_p$ must be
\begin{eqnarray}
\gamma_p=\frac{1}{2\zeta}\left[-\omega^2+\zeta^2+1-\sqrt{\left(\omega^2-\zeta^2-1\right)^2-4\zeta^2}\,\right]\,,
\label{gammap}
\end{eqnarray}
where we took a minus sign in front of the square root term to make $\gamma_p$ converge. Plugging the $\omega$ and $\zeta$ defined in Eq.~(\ref{mij2}) into Eq.~(\ref{gammap}), we get 
\begin{eqnarray}
\gamma_p = i\frac{\sqrt{2}}{\sqrt{\cosh 2\pi p + \cos 2\pi \nu}
 + \sqrt{\cosh 2\pi p + \cos 2\pi \nu +2 }}\,.
\label{gammap2}
\end{eqnarray}
Note that $\gamma_p$ is pure imaginary. For negative $\nu^2$, Eq.~(\ref{gammap2}) is analytic under substitution $\nu\rightarrow\pm i|\nu|$ as we will explain in section \ref{largemass1}. 

If we trace over the $R$ degree of freedom, the density matrix is found to be diagonalized
\begin{eqnarray}
\rho_L ={\rm Tr}_{R}\,|{\rm BD}\rangle\langle{\rm BD}| 
=\left(1-|\gamma_p|^2\,\right)\sum_{n=0}^\infty |\gamma_p |^{2n}\,|n;p\ell m\rangle\langle n;p\ell m|\,,
\label{densitymatrix1} 
\end{eqnarray}
where we used Eq.~(\ref{bogoliubov2}) and defined $|n;p\ell m\rangle=1/\sqrt{n!}\,(c_L^\dagger)^n\,|L'\rangle$.
Notice that we put the normalization factor $1-|\gamma_p|^2$ because
\begin{eqnarray}
\sum_{n=0}^\infty |\gamma_p |^{2n}=\lim_{n\rightarrow\infty}\frac{1-|\gamma_p|^{2n}}{1-|\gamma_p|^2}\xrightarrow{|\gamma_p|<1}\frac{1}{1-|\gamma_p|^2}\,.
\end{eqnarray}
Then, the entanglement entropy as a function of $p$ and $\nu$ is calculated to be
\begin{eqnarray}
S(p,\nu)=-{\rm Tr}\,\rho_L(p)\,{\rm log}\,\rho_L(p)
=-{\rm log}\,\left(1-|\gamma_p|^2\right)
-\frac{|\gamma_p|^2}{1-|\gamma_p|^2}\,{\rm log}\,|\gamma_p|^2\,.
\label{s}
\end{eqnarray}
Note that this formula is derived under the condition $|\gamma_p|<1$. 
 
The quantity $c_6$ that we want to calculate to get the information about the long range correlation of the quantum state is obtained by integrating over $p$ and a volume integral over the hyperboloid,
\begin{eqnarray}
S_{\rm intr}\equiv c_6=\frac{1}{\pi}\int_0^\infty dp\,p^2S(p,\nu)\,.
\label{final}
\end{eqnarray}

Maldacena and Pimentel plot the entanglement entropy, normalized by the conformally invariant case ($\nu=1/2$), $S_{\rm intr}/S_{\nu=1/2}$, as a function of $\nu^2$. (See the red line in Figure \ref{fig2}). They found that the long range entanglement is largest for small mass (positive $\nu^2$) and decays exponentially for large mass (negative $\nu^2$) in \cite{Maldacena:2012xp}.

\subsection{Large mass range}
\label{largemass1}

For large masses ($m>3/2H$), corresponding to negative $\nu^2$, we need to take care of the phase factors in Eq.~(\ref{mij2}). In the case of the Bunch-Davies vacuum, all factors in front of the matrix in Eq.~(\ref{mij1}) after substitution $\nu\rightarrow \pm i|\nu|$ can be equivalently obtained by performing the analytic continuation on Eq.~(\ref{mij2}) which drops all phase factors, that is,
\begin{eqnarray}
-\frac{\Gamma\left(\nu-ip+1/2\right)}{\Gamma\left(\nu+ip+1/2\right)}
\frac{2\,e^{i\pi\nu}}{e^{2\pi p}+e^{2i\pi\nu}}
&\xrightarrow{\nu\rightarrow\pm i|\nu|}&
-\frac{\Gamma\left(\frac{1}{2}+i\left(\pm|\nu|-p\right)\right)}{\Gamma\left(\frac{1}{2}+i\left(\pm|\nu|+p\right)\right)}\frac{2\,e^{\mp\pi|\nu|}}{e^{2\pi p}+e^{\mp2\pi|\nu|}}\\
&=&\frac{\sqrt{2}\,e^{-p\pi}}{\sqrt{\cosh 2\pi p+\cosh 2\pi|\nu|}}\,\,.
\end{eqnarray}
Thus, we can use Eq.~(\ref{gammap2}) for all ranges of masses.

\subsection{Super-curvature modes}

So far, we have considered only continuous spectrum for the eigenvalue $p$. However, it is known that there exists a discrete mode $p=i\left(\nu-1/2\right)$ in the spectrum, the so-called super-curvature mode~\cite{Sasaki:1994yt}. 
We need to worry about it, namely, the super-curvature mode may contribute to the long-range entanglement of a quantum state. 
However, since the super-curvature modes exist with a spacial value of $p=i\left(\nu-1/2\right)$, it is plausible that the integration would not produce a finite measure unless the super-curvature modes behaves as delta-functions due to unnormalizable nature of the super-curvature mode.
 It would be interesting to investigate this more precisely.

\section{Entanglement entropy of $\alpha$-vacua}
\label{sec:alpha}

In the previous section, we reviewed the entanglement entropy in the Bunch-Davies vacuum \cite{Maldacena:2012xp}. Here, we extend the calculation to more general $\alpha$-vacua, which corresponds to a state filled with particles from the point of view of the Bunch-Davies vacuum \cite{Mottola:1984ar, Allen:1985ux}. The $\alpha$-vacua are also de Sitter invariant, so we can use the same mapping in Figure \ref{fig1} that was used to define a simple representation of the two subspaces. We will examine if the long range entanglement is affected by the state in which particles are pair-created in the vacuum.

\subsection{$\alpha$-vacua}

The CPT invariant $\alpha$-vacua can be parametrized by a single positive real parameter $\alpha$. The Bunch-Davies vacuum is realized when $\alpha=0$. The mode function is obtained by the Bogoliubov transformation from the mode function of the Bunch-Davies vacuum in Eq.~(\ref{bdmf}) such as
\begin{eqnarray}
{\cal U}_{\sigma p\ell m}(t,r,\Omega)=\cosh\alpha\,u_{\sigma p\ell m}(t,r,\Omega) + \sinh\alpha\,u_{\sigma p\ell m}^*(t,r,\Omega)\,.
\end{eqnarray}

The relation between the  annihilation operators in the $\alpha$-vacua and the Bunch-Davies vacuum  is also defined by the Bogoliubov transformation
\begin{eqnarray}
d_\sigma=\cosh\alpha\,a_\sigma - \sinh\alpha\,a_\sigma^\dagger\,.
\label{da}
\end{eqnarray}
The definition of  an $\alpha$-vacuum is then simply
\begin{eqnarray}
d_\sigma|\alpha\rangle =0\,.
\label{alphavacua}
\end{eqnarray}
The scalar field in Eq.~(\ref{phi}) is expanded by those mode functions and operators 
\begin{eqnarray}
\hat\phi(t,r,\Omega) &=& \sum_{\sigma,\ell,m} \int dp 
\left[\,d_{\sigma p\ell m}\,{\cal U}_{\sigma p\ell m}(t,r,\Omega)
+d_{\sigma p\ell m}^\dagger\,{\cal U}^{*}_{\sigma p\ell m}(t,r,\Omega)\,\right]\,.
\end{eqnarray}
 
It is helpful to note that the $\alpha$-vacua are directly related to the Bunch-Davies vacuum and the $R$ and $L$-vacua as
\begin{eqnarray}
|\alpha\rangle  
&=& \exp \left[\frac{1}{2} \tanh \alpha \  a_\sigma^\dagger a_\sigma^\dagger \right]  
|\rm{BD} \rangle \nonumber\\
&=&  \exp \left[\frac{1}{2} \tanh \alpha \   \left( \sum_{q=R,L} \left[\,\xi_{q\sigma}^* \, b_q^\dagger +\delta_{q\sigma} \,b_q \,\right] \right)
\left( \sum_{q=R,L} \left[\,\xi_{q\sigma}^* \, b_q^\dagger +\delta_{q\sigma} \,b_q \,\right] \right) 
 \right] 
\nonumber\\
&&\times
\exp\left(\frac{1}{2}\sum_{i,j=R,L}m_{ij}\,b_i^\dagger\, b_j^\dagger\right) |R\rangle|L\rangle\,,
\end{eqnarray}
where we used Eqs.~(\ref{ab}) and (\ref{bogoliubov1}) and pairs of $\sigma$ are summed over. It is well known that the $\alpha$-vacua are nothing but squeezed states. 
Looking at the above formula, we see that the $\alpha$-vacua should create extra correlations across the $R$ and $L$ sub-systems.

\subsection{Bogoliubov transformation to $R$, $L$-vacua}

We calculate the entanglement entropy of $\alpha$-vacua with the setup of the previous subsection. The calculation is completely parallel to that of the Bunch-Davies vacuum but we start with a different set of creation and annihilation operators. 

We first to find a relation between operators of $\alpha$-vacua and ones of $R$ (or $L$) vacua. Plugging Eq.~(\ref{ab}) into Eq.~(\ref{da}), we get the relation
\begin{eqnarray}
d_\sigma =\sum_{q=R,L}\left[\,
\{\cosh\alpha\,\xi_{q\sigma} - \sinh\alpha\,\delta_{q\sigma}\}\,b_q
+\{\cosh\alpha\,\delta_{q\sigma}^*-\sinh\alpha\,\xi_{q\sigma}^*\}\,b_q^\dagger\,\right]\,.
\end{eqnarray}
Comparing this with Eq.~(\ref{ab}), we find that the Bogoliubov transformation for the original $\xi_{q\sigma}$ and $\delta_{q\sigma}^*$ is
\begin{eqnarray}
\xi_{q\sigma}\rightarrow 
\cosh\alpha\,\xi_{q\sigma} - \sinh\alpha\,\delta_{q\sigma}\,,\qquad
\delta_{q\sigma}^*\rightarrow
\cosh\alpha\,\delta_{q\sigma}^*-\sinh\alpha\,\xi_{q\sigma}^*\,.
\label{xidelta2}
\end{eqnarray}
The Bogoliubov transformation between $\alpha$-vacua and $R$ (or $L$) vacua can be found by the consistency of the definition of the $\alpha$-vacua Eq.~(\ref{alphavacua}) of
\begin{eqnarray}
|\,\alpha\,\rangle = \exp\left(\frac{1}{2}\sum_{i,j=R,L}\tilde{m}_{ij}\,b_i^\dagger\,b_j^\dagger\right)
|R\rangle|L\rangle\,,
\label{bogoliubov3}
\end{eqnarray}
provided
\begin{eqnarray}
\tilde{m}_{ij}= -\{\,\cosh\alpha\,\delta_{i\sigma}^* - \sinh\alpha\,\xi_{i\sigma}^*\,\}\{\,\cosh\alpha\,\xi 
- \sinh\alpha\,\delta\,\}_{\sigma\, j}^{-1} \,\,,
\end{eqnarray}
where we used the first equation in Eq.~(\ref{mij1}) and Eq.~(\ref{xidelta2}). Using the expression of $\xi$ and $\delta$ given by Eq.~(\ref{xidelta1}), we get
\begin{eqnarray}
\tilde{m}_{ij}&=&-\frac{\Gamma(\nu-ip+\frac{1}{2})}{\Gamma(\nu+ip+\frac{1}{2})}
\frac{2}{e^{2\pi p}\,(\cosh\alpha-\sinh\alpha\,e^{-2\pi p})^2+e^{2i\pi\nu}\,(\cosh\alpha+\sinh\alpha\,e^{-2i\pi\nu})^2}\nonumber\\
&&\times\left(
\begin{array}{cc}
D_{RR} & D_{RL} \vspace{1mm}\\
D_{LR} & D_{LL} \\
\end{array}
\right)\,,
\label{mij3}
\end{eqnarray}
where
\begin{eqnarray}
D_{RR}=D_{LL}&=&\left(\cosh^2\alpha\,e^{i\pi\nu}+\sinh^2\alpha\,
e^{-i\pi\nu}\right)\cos\pi\nu
-\sinh2\alpha\sinh^2\pi p\,,\\
D_{RL}=D_{LR}&=& i\left[
\cosh^2\alpha\,e^{i\pi\nu}+\sinh^2\alpha\,e^{-i\pi\nu}
+\sinh2\alpha\cos\pi\nu
\right]\sinh\pi p\,.
\end{eqnarray}

\subsection{Diagonalization}

In order to trace out the $R$ (or $L$) degree of freedom, the density matrix has to be diagonalized as in the form Eq.~(\ref{densitymatrix1}). To do this, we need to perform the Bogoliubov transformation Eq.~(\ref{bc}) again to get the relation
\begin{eqnarray}
|\alpha\rangle = \exp\left(\gamma_p\,c_L^\dagger\,c_R^\dagger\,\right)|R'\rangle|L'\rangle\,.
\label{bogoliubov3}
\end{eqnarray}
The consistency conditions for Eq.~(\ref{bogoliubov3})
\begin{eqnarray}
c_R\,|\alpha\rangle= \gamma_p\,c_L^\dagger\,|\alpha\rangle \,,\qquad 
c_L\,|\alpha\rangle = \gamma_p\,c_R^\dagger\,|\alpha\rangle\,,
\end{eqnarray}
gives rise to the system of four homogeneous equations Eqs.~(\ref{system1}) and (\ref{system2}). Here, $\omega$ and $\zeta$ in the case of $\alpha$-vacua are read off from Eq.~(\ref{mij3}) and
\begin{eqnarray}
\omega\equiv \tilde{m}_{RR}=\tilde{m}_{LL}\,,\qquad\zeta\equiv \tilde{m}_{RL}=\tilde{m}_{LR}\,.
\label{omegazeta}
\end{eqnarray}

We see that $\omega$ and $\zeta$ are not real and pure imaginary respectively for positive $\nu^2$, which are different from the case of de Sitter vacuum. Thus we cannot reduce the system of four homogeneous equation Eqs.~(\ref{system1}) and (\ref{system2}) into two by setting $v^*=\bar{v}$ and $u^*=\bar{u}$ in the case of $\alpha$-vacua. We need to solve the system of four homegeneous equations for positive $\nu^2$, with conditions $|u|^2-|v|^2=1$ and $|\bar{u}|^2-|\bar{v}|^2=1$.

Fortunately, we find a non-trivial solution of $\gamma_p$ of this system:
\begin{eqnarray}
 |\gamma_p |^2&=&\frac{1}{2|\zeta|^2}
\left[-\omega^2\zeta^{* 2}-\omega^{*2}\zeta^2+|\omega|^4-2|\omega|^2+1+|\zeta|^4 \right.\nonumber\\
&&\qquad\qquad\left.
-\sqrt{\left(\,\omega^2\zeta^{*2}+\omega^{*2}\zeta^2-|\omega|^4+2|\omega|^2-1
-|\zeta|^4\,\right)^2-4|\zeta|^4}\,\right]\,.
\label{gammap3}
\end{eqnarray}
This recovers Eq.~(\ref{gammap}) when $\alpha=0$, $\omega^*=\omega$ and $\zeta^*=-\zeta$. So these results are analytically consistent. For negative $\nu^2$, we find Eq.~(\ref{gammap3}) is analytic under substitution $\nu\rightarrow -i|\nu|$.\footnote{Here, we have two choices of $\nu\rightarrow\pm i|\nu|$. In the case of the Bunch-Davies vacuum, the result doesn't change whichever sign we choose. In the $\alpha$-vacua case, however, we find that the substitution $i|\nu|$ produces the divergence where $|\gamma_p| \geq 1$. } This can be checked as follows. As we will see in section \ref{largemass2}, $\omega$ becomes real and $\zeta$ becomes pure imaginary for negative $\nu^2$. So we can reduce the system of four equations into that of two equation as in the case of the Bunch-Davies vacuum. We can plot from $\nu^2<0$ by using Eq.~(\ref{gammap}) and we find the plots agree with the ones obtained by using analytic continuation of Eq.~(\ref{gammap3}).

\subsection{Long range entanglement entropy}

\begin{figure}[t]
\vspace{-2.5cm}
\includegraphics[height=11cm]{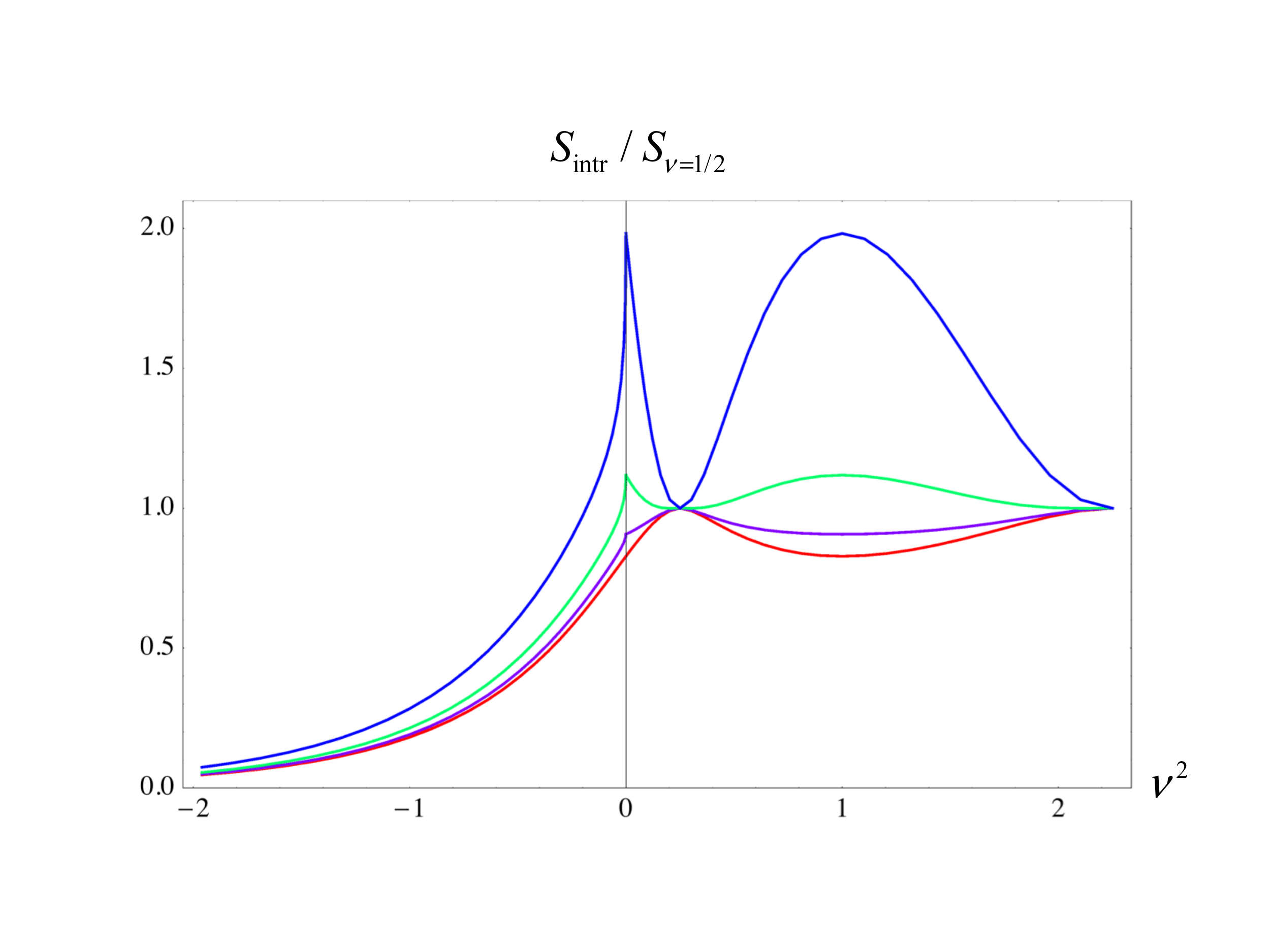}\centering
\vspace{-1cm}
\caption{Plots of the entanglement entropy $S_{\rm intr}/S_{\nu=1/2}$. The red line is for the Bunch-Davies vacuum $\alpha=0$, the purple line is $\alpha=0.03$, the green line is $\alpha=0.1$ and the blue is $\alpha=0.3$. The $S_{\nu=1/2}$ is independent of $\alpha$}
\label{fig2}
\end{figure}

Now, we can plot the entanglement entropy in Eq.~(\ref{final}) for various value of $\alpha$. The results are found in Figure 2.  
The Bunch-Davies vacuum corresponds to $\alpha=0$ (red line). It shows that there exists a long range entanglement for small masses of the scalar field (positive $\nu^2$) and it decays exponentially for large masses (negative $\nu^2$). The entanglement could exist beyond the Hubble horizon because
 de Sitter expansion separates off the pair of particles created within the Hubble horizon. 
 For large masses one expects a lower rate of pair creation and thus a reduced entanglement entropy.

\begin{figure}[t]
\vspace{-2.5cm}
\includegraphics[height=11cm]{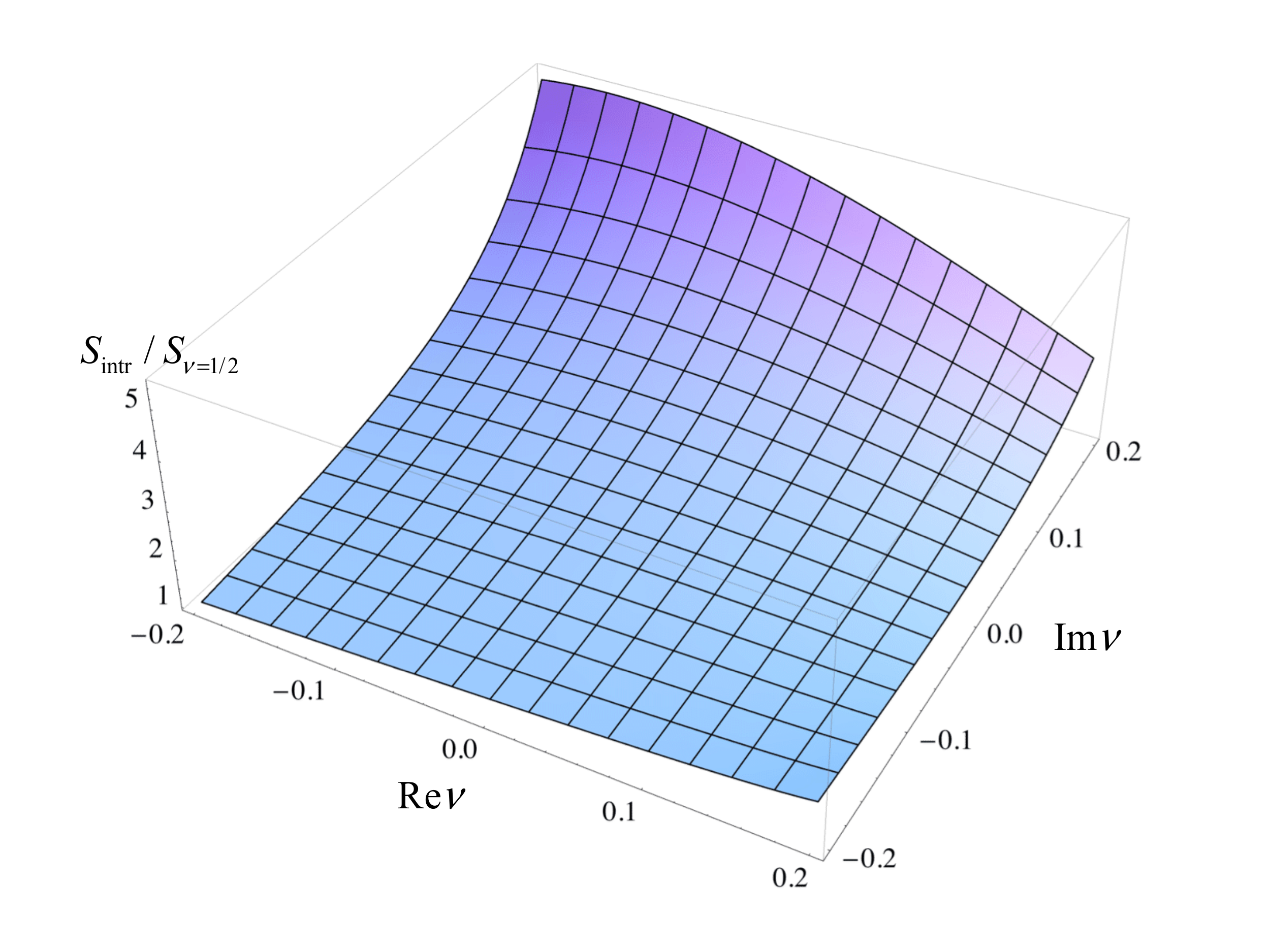}\centering
\vspace{-1cm}
\caption{A plot of the entanglement entropy $S_{\rm intr}/S_{\nu=1/2}$ for $\alpha=0.3$. This is completely smooth around $\nu=0$ in the complex plane.}
\label{fig2-2}
\end{figure}

\begin{figure}[t]
\vspace{-2.5cm}
\includegraphics[height=11cm]{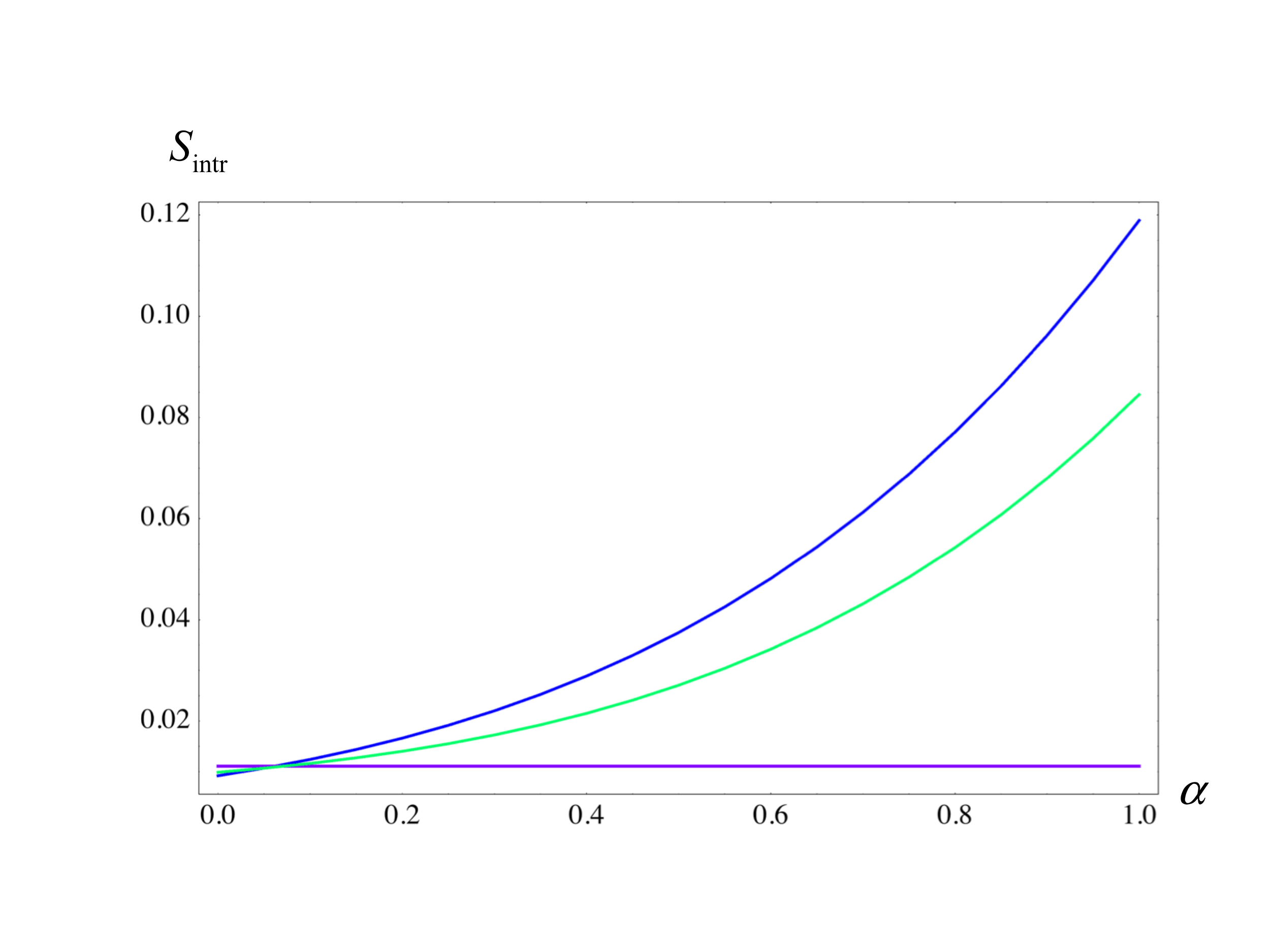}\centering
\vspace{-1cm}
\caption{Plots of the entanglement entropy $S_{\rm intr}$ as a function of $\alpha$. The blue line is for $\nu^2=0$ and $1$, the green line is for $\nu^2=1/16$,\,$9/16$ and $25/16$, and the purple line is for $\nu^2=1/4$ and $9/4$.}
\label{fig2-3}
\end{figure}

The pair condensation effect of the $\alpha$ vacuum in de Sitter space enhances the entanglement entropy, as expected.
Once the effect of a state with condensed particles starts to work as $\alpha$ increases, we see that the entanglement entropy is enhanced, which agrees with our expectation that the state with condensed particles would increase the rate of pair creation by de Sitter expansion for small mass. We also see that the entanglement entropy does not depend on $\alpha$ when the mass parameter $\nu=1/2$. This is because conformally flat space is indistinguishable from Minkowski space for conformally invariant case ($\nu=1/2$). So the differences due to the choice of vacua don't show up at this point. This however is a feature of the particular normalization chosen. For large mass, the long range correlation decreases as the Bunch-Davies vacuum. There appears to be a discontinuity in the derivative in the plots at $\nu^2=0$ as $\alpha$ increases. However, the plots are completely smooth in the complex plane as shown in Figure~\ref{fig2-2}. We also plotted $S_{\rm intr}$ as a function of $\alpha$ for some fixed values of $\nu^2$ in Figure~\ref{fig2-3} where we see the qualitative feature is consistent with Figure~\ref{fig2}.

\subsection{Large mass range}
\label{largemass2}

For the negative $\nu^2$ region ($\nu\rightarrow -i|\nu|$), $\omega$ and $\zeta$ as defined in Eq.~(\ref{omegazeta}) are written by
\begin{eqnarray}
\omega&=&N\left(\cosh 2\alpha\cosh^2\pi |\nu| -\sinh 2\alpha\sinh^2\pi p+\sinh\pi|\nu|\cosh\pi|\nu|\,\right)\,,
\label{omegazeta2}\\
\zeta&=&i\,N\left[\,\left(\cosh 2\alpha +\sinh 2\alpha\right)\cosh\pi|\nu|+\sinh\pi|\nu|\,\right]\sinh\pi p\,,
\end{eqnarray}
where
\begin{eqnarray}
N=-\sqrt{\frac{\cosh (|\nu|-p)}{\cosh (|\nu|+p)}}
\frac{2}{\left(e^{2\pi p}+e^{2\pi|\nu|}\right)\cosh^2\alpha
+\left(e^{-2\pi p}+e^{-2\pi|\nu|}\right)\sinh^2\alpha}\,,
\label{omegazeta3}
\end{eqnarray}
where we have dropped the unimportant phase factor. Note that $\omega$ is real and $\zeta$ is purely imaginary. Thus we can reduce the system of four homogeneous equation Eqs.~(\ref{system1}) and (\ref{system2}) into two Eq.~(\ref{system1}) as we did in the Bunch-Davies case.
We checked that the plots completely agree with each other in the negative $\nu^2$ region.

\subsection{R\'enyi entropy}

\begin{figure}[t]
\vspace{-2.5cm}
\includegraphics[height=11cm]{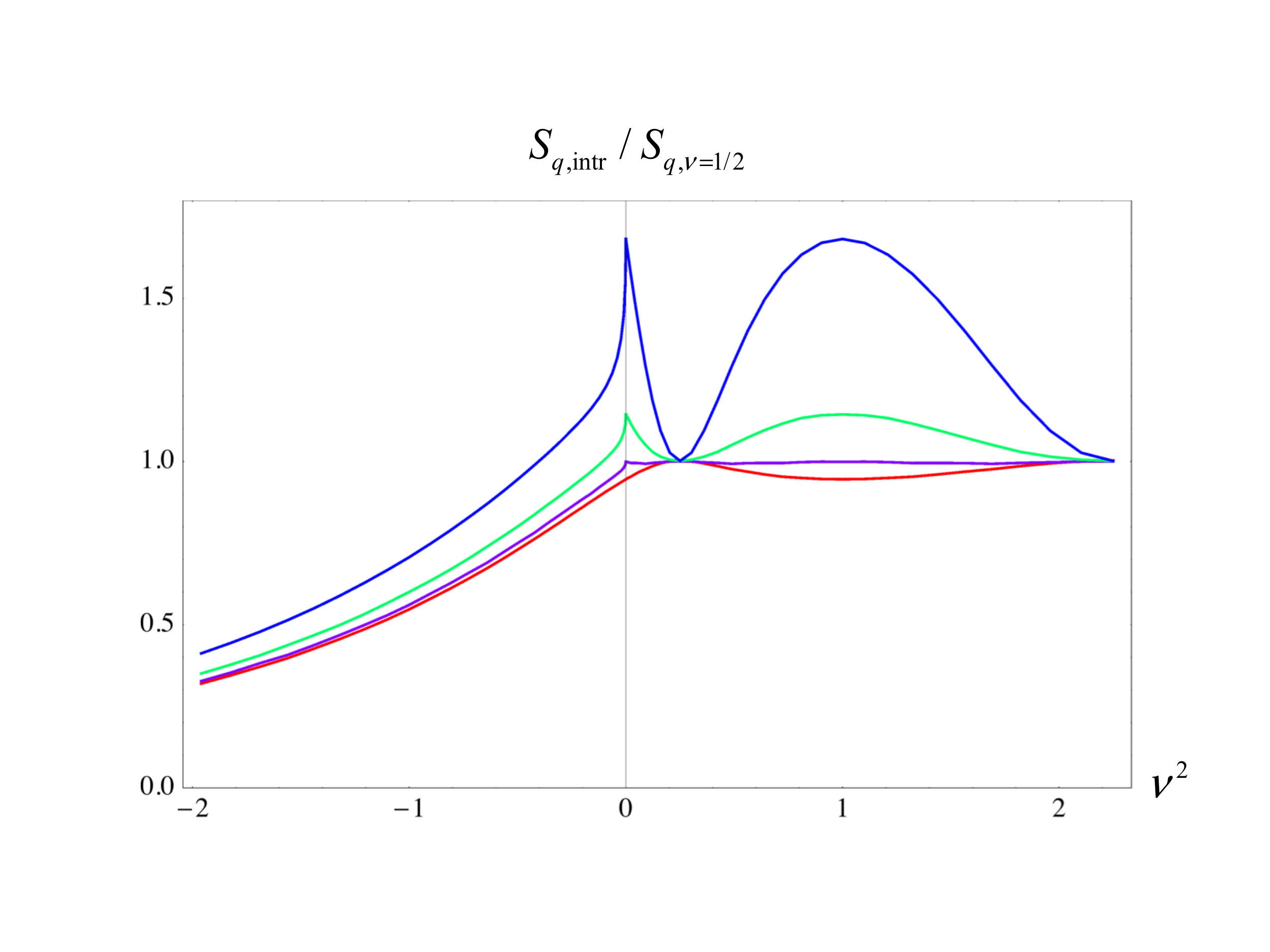}\centering
\vspace{-1cm}
\caption{Plots of the R\'enyi entropy $S_{q,\rm intr}/S_{q,\nu=1/2}$. The order of R\'enyi entropy is $q=1/2$. The red line is for the Bunch-Davies vacuum $\alpha=0$, the purple line is $\alpha=0.03$, the green line is $\alpha=0.1$ and the blue is $\alpha=0.3$.}
\label{fig3}
\end{figure}

The entanglement entropy characterizes the features of a quantum state. However, apparently, it is not a unique such characterization. Actually, there exists a one parameter family generalization of the entanglement entropy,
 the so-called R\'enyi entropy~\cite{Renyi}\cite{Klebanov:2011uf}.
It is defined by
\begin{eqnarray}
S_q=\frac{1}{1-q}\log{\rm Tr}\rho^q\,,\quad q>0\,,
\end{eqnarray}
with the limit $q\rightarrow 1$ corresponding to the entanglement entropy. In fact, this gives rise to a convenient way to calculate the entanglement entropy. In another limit  $q\rightarrow 0$, the R\'enyi entropy measures a dimension of the density matrix, the so-called Hartley entropy~\cite{Headrick:2010zt}.
In the limit  $q\rightarrow \infty$, on the other hand, the R\'enyi entropy tells us the largest eigenvalue of the density matrix.
Thus, the R\'enyi entropy is useful to look at details of a quantum state.

\begin{figure}[t]
\vspace{-2.5cm}
\includegraphics[height=11cm]{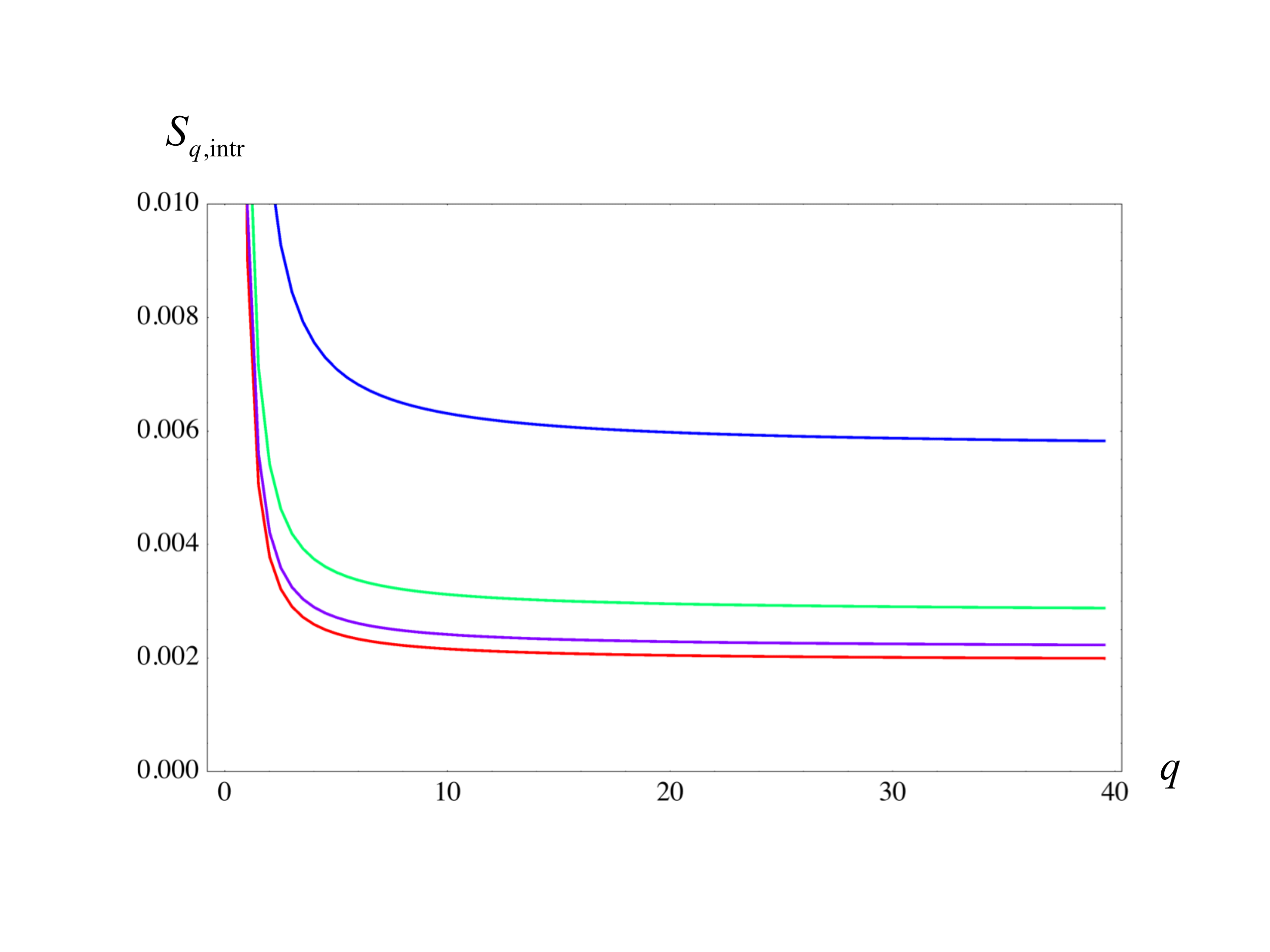}\centering
\vspace{-1cm}
\caption{Plots of the R\'enyi entropy $S_{q,\rm intr}$ as a function of $q$. The mass parameter is $\nu=1$. The red line is for the Bunch-Davies vacuum $\alpha=0$, the purple line is $\alpha=0.03$, the green line is $\alpha=0.1$ and the blue is $\alpha=0.3$.}
\label{fig4}
\end{figure}

In the present cases, the R\'enyi entropy as a function of $p$ and $\nu$ is
\begin{eqnarray}
S_{q,\rm intr}(p,\nu)=\frac{q}{1-q}\log\left(1-|\gamma_p|^2\right)
-\frac{1}{1-q}\log\left(1-|\gamma_p|^{2q}\right) \, ,
\label{renyi}
\end{eqnarray}
where $\gamma_p$ is defined in Eq.~(\ref{gammap3}).
The information about the long range correlation of the quantum state is obtained by integrating over $p$ and a volume integral over the hyperboloid, which is expressed by
\begin{eqnarray}
S_{q,\rm intr}=\frac{1}{\pi}\int_0^\infty dp\,p^2S_q(p,\nu)\,.
\end{eqnarray}
This is plotted in Figure \ref{fig3}. We took the order of the R\'enyi entropy $q=1/2$. 
We see the effect of $\alpha$ enhances the R\'enyi entropy compared with the Bunch-Davies vacuum for small mass. For large mass, the rate of decreasing appears slower than the case of the entanglement entropy. The plots eventually become flat irrespective of $\alpha$ as $q\rightarrow 0$. Remarkably, the R\'enyi entropy exists in spite of the heavy mass. In Figure \ref{fig4}, we plotted $S_{q,\rm intr}$ as a function of $q$. We find 
$S_{q,{\rm intr}}$ is a monotonic function of $q$. It diverges in the limit $q\rightarrow 0$ and approaches the asymptotic value in the limit $q \rightarrow \infty$. The asymptotic value of the R\'enyi entropy increases as $\alpha$ increases. This means the largest eigenvalue of the density matrix is an increasing function of $\alpha$. These features can be explained as follows. From Eq.~(\ref{renyi}), we see 
$ S_{q,\rm intr}(p,\nu) \propto - \log q $ which is independent of $\alpha$ in complete agreement with the numerical result Figure \ref{fig4}
and the fact that Hartley entropy measures the dimensions of the density matrix.
We can also evaluate $S_{q,\rm intr}(p,\nu)$ in the limit  $q \rightarrow \infty$ which results in the formula
\begin{eqnarray}
  S_{q=\infty,\rm intr}(p,\nu) = - \frac{1}{\pi}\int_0^\infty dp\,p^2 \log \left( 1 - \left|  \gamma_p \right|^2   \right)       \   .
\end{eqnarray}
It is easy to evaluate this integral and find that it increases exponentially as $\alpha$ increases.

\section{Discussion}

We studied the entanglement entropy of a free massive scalar filed in de Sitter space by using a method developed by Maldacena and Pimentel. 
In particular, we focused on the state dependence of the entangle entropy, namely, we have considered $\alpha$-vacua which is de Sitter invariant. 
The $\alpha$-vacua can be thought of as a state filled with pair of particles from the point of view of the Bunch-Davies vacuum. 
In order to obtain the entanglement entropy, we need to determine the density matrix of the subsystem $A$ in terms of the notation
 in section 2.1. This can be done when the density matrix is diagonal. To make the density matrix diagonal, it is essential to
transform the original matrix into a special form. This can be done for the Bunch-Davies vacuum. In the case of $\alpha$-vacua
it is not so straightforward to obtain diagonal density matrix. However, we found a formula which can be used to obtain the entanglement entropy.
 We find that the entanglement entropy in de Sitter space takes the minimal value for the Bunch-Davies vacuum among the $\alpha$-vacua.
We argue these features stem from the pair 
condensation of $\alpha$-vacua where the pair has the intrinsic quantum correlation. 
We also calculated the R\'enyi entropy and showed that the asymptotic value of
the R\'enyi entropy increases exponentially as $\alpha$ increases.

In terms of future extensions of this work, one of the most pressing would be a holographic interpretation along the lines of the work of Maldacena and Pimentel for the Bunch-Davies vacuum in \cite{Maldacena:2012xp}. To summarize their argument; a free field theory on a curved de Sitter space 
\begin{eqnarray}
ds^2=dw^2+\sinh^2w\left[-dt^2+\cosh^2t\left(d\theta^2+\cos^2\theta\,d\Omega^2\right)\right]\,.
\label{5ads2}
\end{eqnarray}
can be investigated in a holographic context by treating the de Sitter space as the boundary of an anti-de Sitter (AdS) bulk which can then be considered as the gravitational dual~\cite{Hawking:2000da}\cite{Koyama:2001rf}. 
Here, $d\Omega^2$ is the metric of a two-sphere and $w, t, \theta$ are a radial, a time, and an angular coordinate, respectively.
Then, following Ryu and Takayanagi \cite{Ryu:2006bv}, they match the field theory computation of the entanglement entropy to the area of an appropriate minimal surface in the AdS space. The minimal surface is identified as the surface $t=$ constant and $\theta=0$. 
Starting with a conformally coupled scalar ($\nu = 1/2$),  they generalize the model to non-conformal theories in this holographic scheme. 
Then, the bulk geometry is modified as 
\begin{eqnarray}
  ds^2 = dw^2 + a^2 (w)\, ds_{\rm dS4}^2\,,
\end{eqnarray}
where $ds_{\rm dS4}^2$ is a four-dimensional de Sitter metric and the function $a(w)$ is determined by solving the Einstein equation with the scalar field. 
The surface we should take is the one at $\theta=0$, $t = t(w)$ which extremizes the area  
\begin{eqnarray}
 {\cal A} = V_{S^2} \int a^2 \cosh^2 t(w)\sqrt{dw^2 - a^2 dt^2 }\,,
\label{modify-area}
\end{eqnarray} 
where $V_{S^2}$ is two-volume. 
It would be of great interest to understand how this line of reasoning is modified for $\alpha$-vacua.

The first hurdle to such a holographic computation of the entanglement entropy of the $\alpha$-vacua in a strongly coupled phase is the lack of clarity as to what precisely the gravity duals of a field theory on such $\alpha$ backgrounds are beyond the fact that they will likely be some one-parameter deformation of AdS.
Nevertheless, having demonstrated that at $\nu=1/2$ the entanglement entropy is $\alpha$-independent, we believe that in this case, the Maldacena-Pimentel argument should go through essentially unchanged since the latter does not require a choice of vacuum. More generally though, as we saw above, the entanglement entropy of $\alpha$-vacua is decidedly different from that of the Bunch-Davies vacuum. Hence, we expect the three-dimensional area at $\theta=0$, $t = t(w)$ in the AdS bulk must be further deformed to match the entanglement entropy. 
However, it is not clear how to implement this state dependence in to the holographic scheme\footnote{There is a possible approach to this direction in \cite{Fischler:2013fba}}. In the conventional case, the boundary geometry is Minkowski space where we do not have vacuum ambiguity. Now that there is a continuum of vacua in de Sitter space, we are not sure what surface corresponds to each vacuum. A naive answer is that because there is a difficulty in defining an interacting field theory on an $\alpha$-vacuum ~\cite{Danielsson:2002mb,Einhorn:2003xb,Collins:2003zv} the holographic principle may not lead to a well-defined gravitational theory. We leave these considerations for future investigation.

\section*{Acknowledgments}
We would very much like to thank Juan Maldacena and Guilherme Pimentel for their invaluable advice and assistance during the writing of this paper. This work was supported in part by funding from the University Research Council of the
University of Cape Town, Grants-in-Aid for Scientific Research (C) No.25400251 and Grants-in-Aid for Scientific Research on Innovative Areas No.26104708. JM is supported by the National Research Foundation of South Africa through the IPRR and CPRR programs.

%

\end{document}